\documentclass{ws-procs9x6}
\usepackage{bbm}

\begin{document}

\title{Nonperturbative access to Casimir-Polder forces}

\author{B. D\"OBRICH$^{1,*}$and H. GIES$^{1,2}$}

\address{$^1$Theoretisch-Physikalisches Institut, Friedrich-Schiller-Universit\"at Jena,\\
 $^2$Helmholtz Institute Jena, Max-Wien-Platz 1,D-07743 Jena, Germany\\
$^*$E-mail: babette.doebrich@uni-jena.de}

\author{M. DEKIEVIET}

\address{Physikalisches Institut,
    Universit\"at Heidelberg, \\ Philosophenweg 12, D-69120 Heidelberg, Germany}

\begin{abstract}
We discuss the scalar analogue of the Casimir-Polder force between a sphere and a uniaxially corrugated surface with Dirichlet boundary conditions. Presenting a formulation that is nonperturbative in the height profile of the surface, we give explicit numerical results for a sinuosoidal corrugation profile.
\end{abstract}


\bodymatter

\section{Introduction}

Past years have witnessed great progress in the study of Casimir-Polder~\cite{CasimirPolder} forces between an atom and a surface. On the theoretical side, the dependence of the Casimir-Polder force on the surface geometry is an important problem~\cite{bezerra,messina,Dalvit:2008zz,Buhmann:2006ji}. Often, its deviation from the standard planar situation is accounted for in a perturbative manner: the (mean) amplitude $A$ of the surface corrugation is assumed to be the smallest length scale of the system. However, in recent high resolution experiments using the atomic beam spin echo technique~\cite{DeKieviet}, the atom-wall distance can become much smaller than the amplitude of the surface corrugation. 
Thus, there is an urgent need for nonperturbative calculations of the Casimir-Polder potential. Starting with the scalar Dirichlet situation, we present such a treatment in the following.

\section{Nonperturbative access to scalar fields}

The presence of bodies or surfaces imposes boundary conditions on fluctuating quantum fields. This gives rise to a shift in the energy of the ground state, the Casimir energy. A substraction of the Casimir self-energy of the bodies then yields the Casimir interaction energy between the surfaces which serves as a potential energy for the Casimir force. Using the constrained-functional integral approach~\cite{Bordag:1983zk,Emig:2001dx}, the boundaries on the fluctuating field are implemented through a $\delta$ functional. Upon integration over the fields, the Casimir interaction energy between two surfaces $S_{1}$ and $S_{2}$, separated by a (mean) distance $H$ can be written as
\begin{equation}
E(H) = -\frac{\hbar
  c}{T_{\mathrm{E}}}\frac{1}{2}\sum_{n=1}^{\infty}\frac{1}{n}
\mathsf{Tr}\left(\mathcal{M}_{11}^{-1}\mathcal{M}_{12}
\mathcal{M}_{22}^{-1}\mathcal{M}_{21}\right)^{n}\ ,\label{E_tr_cycle} 
\end{equation}
where $T_{\mathrm{E}}$ denotes the length in Euclidean time direction.
$\mathcal{M}_{\alpha\beta}$ is the propagator of the fluctuations, i.e.
\begin{equation}
\mathcal{M}_{\alpha\beta}(\zeta,\vec{x}-\vec{x}')=\frac{1}{4\pi|\vec{x}-\vec{x}'|}\exp{(-|\vec{x}-\vec{x}'| |\zeta|)} \label{MAlphaBeta}
\end{equation}
for the scalar Dirichlet case.

In Eq.~\eqref{MAlphaBeta}, $\zeta$ denotes the imaginary frequency, while $\vec{x}=(x_1,x_2,x_3)$ and $\vec{x}'$ are three-vectors pointing onto the surfaces $S_{\alpha}$ and $S_{\beta}$, respectively. As the surfaces respond to the field by charge fluctuations, the inverse propagator $\mathcal{M}_{\alpha\beta}^{-1}$ can be interpreted as the propagator of charge fluctuations within the surface. The trace in Eq.\eqref{E_tr_cycle} has to be taken over the coordinates of the surfaces, demanding the inclusion of appropriate metric factors for the integration measures. Furthermore, the functional inverse of $\mathcal{M}_{\alpha\beta}$ is generally not analytically known for nontrivial surfaces.

In the following, we evaluate the Casimir energy between a surface $S_{1}$ which is uniaxially corrugated along the direction $x_1$ and a sphere $S_{2}$ with radius $r$, cf. left panel of Fig.~\ref{aba:fig1}. We are interested in the Casimir-Polder limit ($r\ll H$), where the analytical result for a flat surface $S_{1}$ is known to be  $\mathcal{O}(\frac{r}{H^2})$ to leading order~\cite{Bulgac:2005ku}.

We compute $\mathcal{M}_{22}^{-1}$ from $\mathcal{M}_{22}^{-1} \mathcal{M}_{22} = \mathbbm{1}$, where $\mathcal{M}_{22}$ is given through~\eqref{MAlphaBeta}. By expansion of the equation in terms of spherical harmonics $Y_{lm}$, $\mathcal{M}_{22}^{-1} $ can be calculated to arbitrary order in $l$. For the computation of the leading order Casimir-Polder energy, however, it suffices to consider the monopole contribution $l=0=m$, which reads $\mathcal{M}_{22}^{-1}(\zeta)=\left|\zeta\right|\exp(r|\zeta|)/\left[4\pi r^{2}\sinh(r|\zeta|)\right]$.

Next, we go over to dimensionless variables
by a rescaling with the distance parameter $H$: $\vec{x}\rightarrow \tilde{\vec{x}} H$, $\zeta \rightarrow \tilde{\zeta}/H$. We find that to first order $\mathcal{O}(\frac{r}{H^2})$ in the Casimir energy, it is sufficient to consider the $n=1$ term of the sum in Eq.~\eqref{E_tr_cycle}. Furthermore, in this limit the propagators $\mathcal{M}_{12}$ and $\mathcal{M}_{21}$ become independent of the coordinates on the sphere $S_2$. As the monopole order of $\mathcal{M}_{22}^{-1}$ is also independent of these coordinates, the two integrations over the surface of the sphere contribute only a factor of $16\pi^2$ in Eq.~\eqref{E_tr_cycle}. Only after this step, the translational invariance of the surface $S_1$ along the 2-component can be exploited by a Fourier transformation of Eq.~\eqref{E_tr_cycle} to momentum space.

In a final step, substituting $\tilde{q}=\sqrt{\tilde{\zeta}^2+\tilde{p}_2^2}$, Eq.~\eqref{E_tr_cycle} reduces to:
\begin{equation}
E=-\frac{\hbar c r}{H^2}{\int_{0}^{\infty}\mathrm{d}\tilde{
  q}\int_{-\infty}^{\infty}\mathrm{d}\tilde x\,\sqrt{g(\tilde x)}\tilde q
\Delta\tilde{\mathcal{M}}_{12}(\tilde q;\tilde
x)\tilde{\mathcal{M}}_{21}(\tilde q;\tilde x)}
+\mathcal{O}\left(\frac{r^{2}}{H^{3}}\right), \label{E_final}
\end{equation}
where we have defined
$\Delta\tilde{\mathcal{M}}_{12}=\tilde{\mathcal{M}}_{11}^{-1}\tilde{\mathcal{M}}_{12}$ and dropped the coordinate subscript ''1'':
$\tilde{x}_1\rightarrow \tilde{x}$. The metric factor is related to the height profile $h(\tilde{x})$ by
\begin{equation}
\sqrt{g(\tilde x)}=\sqrt{1+\left(\partial_{\tilde x} \tilde h(\tilde 
x )\right)^{2}},
\quad \tilde h(\tilde x) = \frac{1}{H}\, h( \tilde{x} H)\ .
\label{metric_resc_general_height}
\end{equation}

In principle, computing the energy in Eq.~\eqref{E_final} is now very simple. The combined propagator $\Delta\mathcal{M}_{12}$ can be obtained by solving
\begin{equation}
\int_{\tilde x}\sqrt{g(\tilde x)}\tilde{\mathcal{M}}_{11}(\tilde
q;\tilde{x}';\tilde x)
\Delta\tilde{\mathcal{M}}_{12}(\tilde q;\tilde x)
=\tilde{\mathcal{M}}_{12}(\tilde q;\tilde x') 
\label{bestimmungsgl_deltaM12_final}
\end{equation}
numerically.
However, the treatment of the above equation is nontrivial due to the singular structure of $\mathcal{M}_{11}$ at the origin, see Eq. \eqref{m11_resc_general_height}. Thus, a suitable regularization scheme has been worked out~\cite{Dobrich:2008zz}.

The dimensionless propagators $\tilde{\mathcal{M}}_{12}\equiv
\tilde{\mathcal{M}}_{21}$ and $\tilde{\mathcal{M}}_{11}$ that enter Eqs. \eqref{bestimmungsgl_deltaM12_final} and \eqref{E_final} are given in terms of Bessel functions:
\begin{eqnarray}
\tilde{\mathcal{M}}_{11}(\tilde q;\tilde{x}';\tilde x) & = &
\frac{1}{2\pi}K_{0}
\left(\tilde q\sqrt{(\tilde x'-\tilde x)^{2}+\left(\tilde{h}(\tilde{x}')-\tilde{
      h}(\tilde x )\right)^{2}}\right) , \label{m11_resc_general_height}\\
\tilde{\mathcal{M}}_{12}(\tilde q;\tilde x') & = & \frac{1}{2\pi}K_{0}
\left(\tilde q\sqrt{(\tilde x')^{2}+\left(\tilde h(\tilde{
      x}')-1\right)^{2}}\right)\ .
\label{m12_resc_general_height}
\end{eqnarray}

\section{Results for a sinusoidal surface corrugation}

As a concrete example, we consider the case of a sinusoidal surface corrugation. To this end, we employ  $h(x)=A\sin(\omega x+\phi)$ as height function in Eq.~\eqref{metric_resc_general_height}. We fix the center of the sphere at $x=0$ and use the phase $\phi$ to effectively vary the sphere's position above the structure. 

\begin{figure}
\begin{minipage}{0.4 \linewidth}
\psfig{file=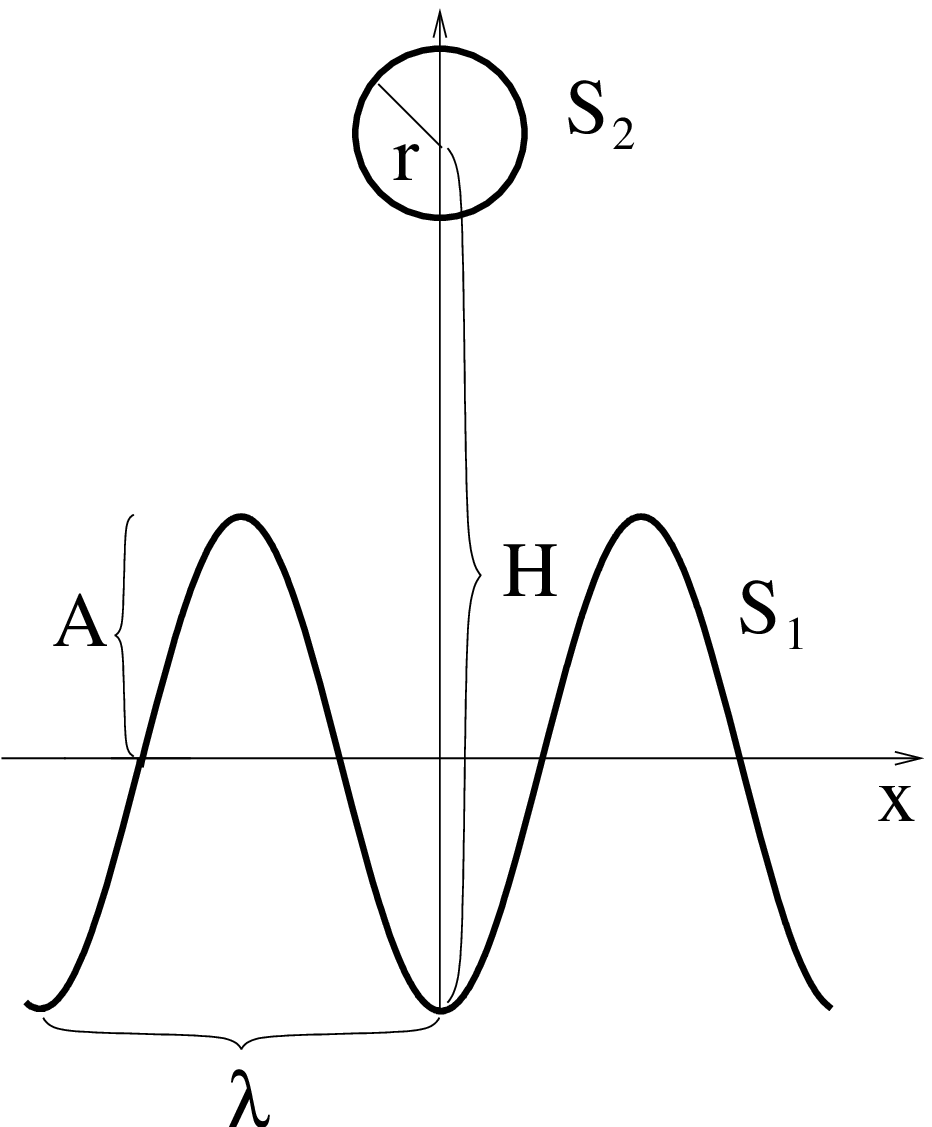,scale=0.4}
\end{minipage}
\begin{minipage}{0.58 \linewidth}
\psfig{file=largerHfitSineWell.eps,scale=0.45}
\end{minipage}
\caption{Left: Involved length scales of the setup: Sphere $S_2$ of radius $r$ at distance $H$ above a corrugated surface $S_1$ with amplitude $A$ and corrugation wavelength $\lambda=\frac{2\pi}{\omega}$. Right: $E_{\text{sine}}/E_{\text{planar}}$ as a function of separation $H/A$ for corrugation frequencies $\omega A=1,2,3$}
\label{aba:fig1}
\end{figure}

In the following, we present numerical results for the energy above a minimum of the structure, i.e. for $\phi=-\frac{\pi}{2}$. In order to highlight the geometry-induced effects, we normalize the results for the Casimir energy $E_{\text{sine}}$ with respect to the energy of the planar-surface situation $E_{\text{planar}}$. For consistency, $E_{\text{planar}}$ is also evaluated numerically.
On the right panel of Fig.~\ref{aba:fig1}, we display $E_{\text{sine}}/E_{\text{planar}}$ as a function of the normalized distance $H/A$ for three different corrugation frequencies $\omega A=1,2,3$. In the limiting cases of $H/A \rightarrow \infty$ and $H/A \rightarrow 0$, we find that $E_{\text{sine}}/E_{\text{planar}}\rightarrow 1$, as can be expected: For $H/A \rightarrow \infty$, the corrugation cannot be resolved by the sphere as it is much smaller than the distance, whereas for $H/A \rightarrow 0$, the corrugation is much larger than the distance and is thus not seen locally. By contrast, for distances $H\sim A$, a distinct deviation from the planar-surface case is found. As the deformation of the surface at the structure minimum ''bends'' towards the sphere, one finds $E_{\text{sine}}/E_{\text{planar}}> 1$. This effect becomes more pronounced as the structure wells become more narrow, i.e. for larger $\omega A$.

It is useful to parameterize the deviation of the Casimir energy in the non-planar situation from the flat-surface setup in terms of an anomalous dimension $\eta$, by defining $E_{\text{corrugation}}\sim 1/H^{2+\eta}$, where $\eta=0$ for a flat surface $S_1$. The increase of the normalized Casimir-Polder energy towards the peak is found to scale linearly, corresponding to $\eta=-1$, with $\omega$-dependent linear coefficients (not shown in Fig.~\ref{aba:fig1}). In the drop-off region of the potential right beyond the peak, the anomalous dimension depends on $\omega$: For $\omega A=1,2,3$ we find $\eta\simeq 0.4, 1.0,
1.6$, respectively. Most interestingly, at larger distances $H/A\simeq 10$, all curves converge towards a universal curve characterized by an anomalous dimension of $\eta\simeq 0.2$, irrespectively of the frequency $\omega A$.

Within the worldline picture of quantum field theory~\cite{Gies:2003cv} this can be attributed to the fact that the quantum vacuum fluctuations average over the surface geometry as they are isotropic in space. With growing separation between sphere and plate, the effect of higher corrugation frequencies $H\omega\gg 1$ is not resolved anymore, as has been also confirmed by studies of a sawtooth-like corrugation~\cite{Dobrich:2008zz}.

\section{Conclusions}

In this work we have investigated the scalar analogue of Casimir-Polder energies between a sphere and a uniaxially corrugated surface, using a sinusoidal surface profile as an example. In particular, our study was not based on a perturbative ordering of length scales and thus allows for arbitrary ratios of the objects' separation $H$ and the deformation parameters $\omega$ and $A$ in the limit of vanishing sphere radius $r$.
In a numerical study we have parameterized the geometry-dependence of the Casimir energy by introducing an anomalous dimension $\eta$, which was shown to be non-integer valued in the regime of $H\sim A$. This result is not accessible through a perturbative calculation. Although our results for the Dirichlet scalar case should not be viewed as a quantitative estimate for the electromagnetic case, we expect analogous results for the anomalous dimensions also for the latter case. This is currently under investigation.

\section*{Acknowledgments}
B.D. would like to thank the organizers of QFEXT 09 for the opportunity to present this work. Financial support through DFG/TR18, DFG/GRK1523 and DFG/Gi328/5-1 is greatfully acknowledged.

\end{document}